\documentclass[12pt]{iopart}
\usepackage{graphicx}
\begin{document}
\jl{1}
\title[Comment]{Reply to Comment on\\
'A new method to calculate the spin-glass order parameter
of the two-dimensional $\pm J$ Ising
model'}

\author{Hidetsugu 
Kitatani\footnote[3]{E-mail address:kitatani@vos.nagaokaut.ac.jp}
 and Akira Sinada}

\address{Department of Electrical Engineering, Nagaoka University of
Technology, Nagaoka, Niigata 940-2188, Japan}

\begin{abstract}
In response to the comment made by Dr. Shirakura {\it et al}
(cond-mat/0011235),
we explain that their scaling forms of the
order parameter distribution are inadequate.
We  then present  an appropriate scaling form 
of the order parameter distribution,
which gives a good scaling plot of the order parameter distribution 
itself 
and also gives  consistent results  with our
pevious results[1]. 
\end{abstract}

%
%

In a recent paper[1], we dicussed the spin-glass phase transition
in the two-dimensional $\pm J$ Ising model.
We analysed the Binder parameter, $g_{L}$, and the spin-glass 
susceptibility, $\chi_{\rm SG}$, 
by finite-size scaling including the corrections to scaling.
We  concluded that the estimation of $T_{\rm c}$ is strongly
affected by the corrections to scaling, so that there still remains
the possibility that $T_{\rm c}=0$.
In the comment by Shirakura {\it et al}[2], they have performed the
scaling analysis of the spin-glass order parameter distribution, 
$P_{L}(q)$.
They insisted that two scaling forms of $P_{L}(q)$ 
reproduce the scaling forms of $g_{L}$ and $\chi_{\rm SG}$ in [1], 
but do not 
give good scaling plots of the order parameter distribution itself.
Thus they concluded that there is no possibility that  $T_{\rm c}=0$.

In this reply, we first explain  that the scaling forms of $P_{L}(q)$
proposed in [2]
are inadequate. Then, we present
an appropriate scaling form of $P_{L}(q)$,
which gives
a good scaling plot of $P_{L}(q)$ itself, and gives
consistent results with our previous results[1].

First, the integral of rhs of equation (1)(equation (2)) in [2]
gives $1/(1+c/L^{\omega})$ ($(1+d/L^{\omega})^{1/2}$).
Namely, both the scaling forms of $P_{L}(q)$ 
proposed in [2] are not normalized correctly.
This should not be ignored, since
the difference from 1 is $O(1/L^{\omega})$, which is the
same order with the difference of the ratios of  peak heights 
for various $L$ in the scaling
plots of $P_{L}(q)$.  
Furthermore, when the scaling forms of 
 $P_{L}(q)$ (equations (1) and (2)) in [2] are correctly normalized,
we obtain that $a=0$ and $b=2c=-d$, where $a$ and $b$ are 
the coefficients
of the correction terms in the scaling forms of
 $g_{L}$ and $\chi_{\rm SG}$ (equations (16) and (17))
in [1]. 
Namely, the scaling forms of $P_{L}(q)$ proposed in [2]
give no correction term of $O(1/L^{\omega})$ in
the scaling form of  $g_{L}$ for any values
of $c$ and $d$, when they are appropriately normalized.
Thus, we conclude that the scaling forms
proposed in [2] are inadequate.

 We did not consider the scaling form of $P_{L}(q)$ in [1], since
our method only gives $[<q^{n}>_{L,T}]_{p}$.
In principle, however, the correction term of $[<q^{n}>_{L,T}]_{p}$
for any $n$ comes from the correction term of  $P_{L}(q)$.
Thus, from now on, we discuss the  scaling form of
$P_{L}(q)$ including the corrections to scaling at $T=T_{\rm c}$. 

We think that an appropriate scaling form of $P_{L}(q)$ 
at $T=T_{\rm c}$ can
be written as follows:
\begin{equation}
  P_{L}(q) = L^{\eta /2}\bar{P}(qL^{\eta /2})(1+
\frac {\bar{f}(qL^{\eta /2})-k}{L^{\omega}}),
\end{equation}
where $\bar{P}(x)$ and $\bar{f}(x)$ are some scaling functions.
$P_{L}(q)$ is normalized when we take $k$ as
\begin{equation}
  k = \int {\bar{P}(x)\bar{f}(x)}dx.
\end{equation}
Now, we show that a simple form of $\bar{f}(x)$ 
 gives a good scaling plot of $P_{L}(q)$ itself, and gives consistent 
results with
our previous results[1].
We take $\bar{f}(x)$ as follows:
\begin{equation}
\label{cases}
   \bar{f}(x)=\cases{11-10\mid x\mid &for $\mid x \mid \leq 1.1$\\
         0&for $1.1< \mid x \mid .$\\}
\end{equation}

First, we show the scaling plot of $P_{L}(q)$ 
with $\omega=0.5$, $k=1.252$ and $\eta=0.2$[4] in figure 1, where
we can see that the data with different $L$ fit very well on one
scaling function. (Here,  we have used the data of 
$P_{L}(q)$ used in [2] and [3].)
This $\bar{P}(x)$ gives $\bar{g}(T=0)=0.938$ and $a=-0.319$, 
which are consistent with our previous results (figure 7 in [1]).
However, it gives $\bar{\chi}_{\rm SG} (T=0)=1.07$ and $b=-0.729$,
 which are different from our previous results (figure 8 in [1])[5].
Here, we show a scaling plot of $\chi_{\rm SG}$ with
$\omega ' =0.5$, $b=-0.729$, 
$\nu =2.6$, $\eta=0.2$ and $T_{\rm C}=0$ in figure 2. We find that 
the data fit rather well on one scaling function,
though the data with small linear sizes deviate from
the scaling function.
Further,  $\bar{\chi}_{\rm SG} (T=0)=1.07$ is also consistent 
with that in figure 2.
Thus, we conclude that the value of $b$ in equation (17) in [1]
should be modified.
 It is noted that this modification does not 
change the main result in [1] that the estimation of $T_{\rm c}$ 
is strongly
affected by the corrections to scaling, so that there still remains
the possibility that $T_{\rm c}=0$.

In conclusion, we have shown that the scaling forms of the
order parameter distribution, $P_{L}(q)$, 
proposed in [2] are inadequate. Further, we propose an appropriate 
scaling form
of $P_{L}(q)$. A simple example of $\bar{f}(x)$ gives
a nice scaling plot of $P_{L}(q)$, and also gives consistent results
with those in [1], though we have had to modify
the value of the coefficient, $b$, in the scaling
form of the spin-glass susceptibility, $\chi_{\rm SG}$.

\ack
The authors would thank Dr. Shirakura for sending the detailed data
in [2] and [3].

\begin{figure}[htbp]
\begin{center}
\includegraphics[width=120mm]{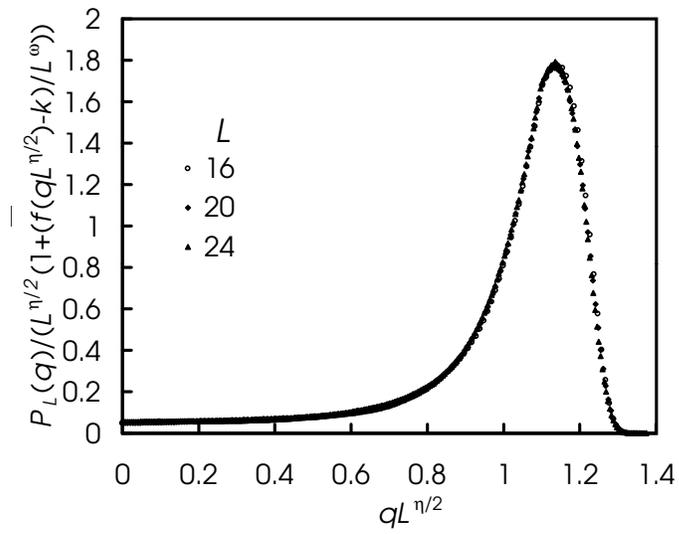}
\end{center}
\caption{A scaling plot of the order parameter distribution, $P_{L}(q)$,
with $\omega=0.5$, $k=1.252$ and $\eta=0.2$,
using the scaling form of equation (1).}
\end{figure}

\begin{figure}[htbp]
\begin{center}
\includegraphics[width=120mm]{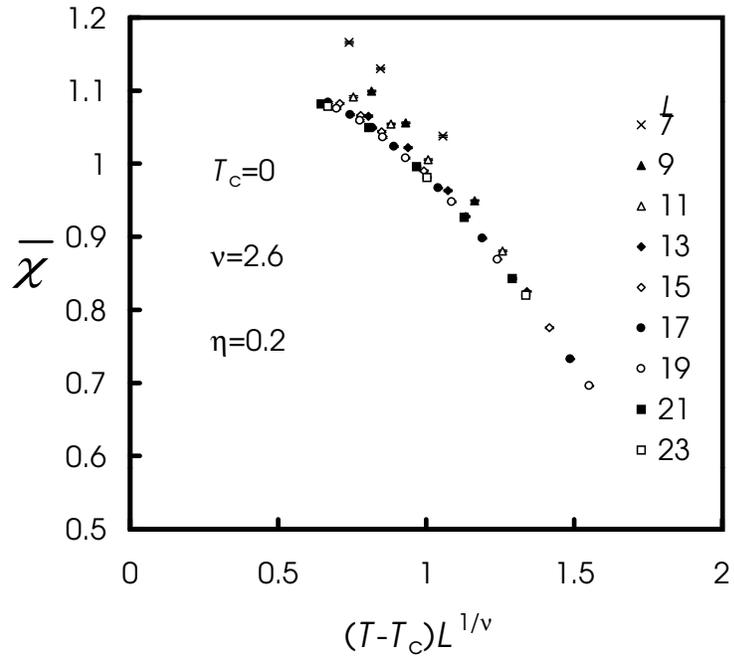}
\end{center}
\caption{A scaling plot of $\chi_{\rm SG}$  
with $\omega ' =0.5$, $b=-0.729$,
$\nu =2.6$, $\eta=0.2$ and $T_{\rm C}=0$, 
using the scaling form of equation (17) in [1].}
\end{figure}


\begin{thebibliography}{100}
\bibitem{note}It is noted that slightly different boundary 
conditions have been used
in [1] and [3]. Therefore, $P_{L}(q)$ may be different in both studies, 
which may lead to different values of 
$a$, $b$ and $\omega$. However,
the values of $\bar{g}(T=0)$ and $\bar{\chi}_{\rm SG}(T=0)$
should be free from the boundary condition.
\end{thebibliography}
\end{document}